\documentclass[11pt,superscriptaddress,aps,prd,preprint]{revtex4}
\usepackage{amsmath}

\makeatletter
\usepackage[T1]{fontenc}
\usepackage{amsmath}
\usepackage{amssymb}
\usepackage{graphicx,color}

\usepackage{slashed}

\newcommand{\bea}{\begin{eqnarray}}
\newcommand{\eea}{\end{eqnarray}}

\begin{document}

\title{G\"{o}del-type Solutions in Cubic Galileon Gravity}

\author{J. R. Nascimento, A. Yu. Petrov, P. Porf\'{\i}rio}
\affiliation{Departamento de F\'{\i}sica, Universidade Federal da Para\'{\i}ba\\
 Caixa Postal 5008, 58051-970, Jo\~ao Pessoa, Para\'{\i}ba, Brazil}
\email{jroberto, petrov, pporfirio@fisica.ufpb.br}

\author{A. F. Santos}
\affiliation{Instituto de F\'{\i}sica, Universidade Federal de Mato Grosso,\\
78060-900, Cuiab\'{a}, Mato Grosso, Brazil}
\email{alesandroferreira@fisica.ufmt.br}

\begin{abstract}

We address the homogeneous in space and time (ST-homogeneous) G\"{o}del-type metrics within the cubic galileon theory, a particular class of generalized galileon theories. We check the consistency of such spacetimes for a physically well-motivated matter content, namely, a perfect fluid and an electromagnetic field. In this scenario, we find that the admissible solutions impose constraints on the constant couplings ($c_{i}$'s) of the cubic galileon theory to ensure the consistency. Also, we show the existence of a vacuum completely causal solution.

\end{abstract}

\maketitle

\section{Introduction}

The General Relativity (GR) is up to now the most successful theory of gravity from the phenomenological point of view, it has been confirmed by highly accurate  experiments both in weak and strong field regimes, see f.e. \cite{LIGO,LIGO2,LIGO3,Akiyama}. In spite of that, the notorious problem of GR is the presence of pathological solutions suffering from unavoidable theoretical problems. For example, Schwarzschild solution suffers from a physical singularity located at $r=0$, that means that the geodesics of particles end at this point and cannot be extended beyond (geodesic incompleteness). Physically speaking, it leads to the impossibility to make measurements of observables at this point and then Einstein equations break down. The further example is the G\"{o}del metric \cite{Godel} that displays another kind of severe conceptual problem -- causality violation. In fact, this metric is plagued by Closed Time-Like Curves (CTC's) that allows an observer traveling along them come back to the past, thus breaking the causality and violating the chronology protection conjecture \cite{Hawking:1991nk}. Such aforementioned questions raised a suggestion that a new approach for gravity should be taken into account, for example, a consistent quantum theory of gravity in which string theory is the most prominent candidate. Alternatively, a promising way also would be consider alternative theories of gravity mainly driven by astrophysical and cosmological observations \cite{Riess}.

As it is well known, there are two main theoretical motivations for developing alternative gravity models -- first, GR displays essential problems at the perturbative level being non-renormalizable, second, it does not succeed to explain accelerated expansion of the Universe (for a review of various manners to implement modifications in gravity see f.e. \cite{mybook}). The main directions of extending GR are, first, modification of the purely gravitational sector through adding higher-derivative terms, second, introducing new fields, usually scalar or vector ones, which have nothing to do with the usual matter, thereby being an extra ingredient to participate in the dynamics. 

The Brans-Dicke theory \cite{Brans:1961sx} has been the first scalar-tensor theory proposed, since then such extended GR models have received increasingly more attention. Recently, another sort of scalar-tensor model called Horndeski theory \cite{Hornd} (see \cite{Kobayashi:2019hrl} for a recent review) have attracted much attention, this model exhibits all possible (non)-minimal couplings between the scalar field and the curvature engendering second-order equations of motion, thus avoiding Ostrogradsky instabilities \cite{Ostro}. The Horndeski theory is equivalent to the generalized covariant galileon theory in four dimensions \cite{Deffayet:2009wt, Deffayet:2009mn, Deffayet:2011gz} -- the mapping between both was first checked out in \cite{Kobayashi:2011nu}. Particularly, the galileon theory was first introduced in flat space \cite{Ratt} in order to study the accelerated expansion of the Universe without need of the cosmological constant. Such a model has a symmetry resembling the Galilean symmetry in mechanics, i.e., $\pi\rightarrow\pi+b_{\mu}x^{\mu}+c$, where $\pi$ is the scalar field commonly called galileon. Conversely, its covariant version, covariant galileon theory, does not possess this symmetry since the Galilean symmetry is broken along the ``covariantization'' procedure \cite{Deffayet:2009wt}. Particular covariant galileon theories have been derived from other contexts, for example: in disformally coupled theories \cite{Zuma}, in the decoupling limit of the Dvali-Gabadadze-Porrati (DGP) braneworld model \cite{Dvali:2000hr} and in the higher-dimensional Einstein-Maxwell Gauss-Bonnet theory after a consistent Kaluza-Klein (KK) reduction to four dimensions \cite{Charmousis:2012dw}.  Also, it is interesting to note a link between galileons and massive gravity \cite{Hint}. However, up to now, most part of papers on galileon gravity is devoted to cosmological studies (see f.e. \cite{KoivMota,Zuma}).

At the same time, besides of cosmological Friedmann-Robertson-Walker (FRW) metric and spherically symmetric metrics (the black holes solutions have been extensively discussed within galileons framework f.e. in \cite{Rinaldi}), one more relatively simple metric deserves to be studied, that is, the G\"{o}del metric which is one of the first known metric displaying causality violation \cite{Godel}. This violation arises due to the existence of CTC's. The further generalization of this metric has been performed in \cite{Reb} where the class of G\"{o}del-type metrics has been defined (various aspects of such metrics have been discussed further in \cite{Reb1}). It is worth emphasizing that G\"{o}del-type metrics have a completely causal region (without CTC's) for a specific relationship between their two parameters distinguishing to G\"{o}del metric. Studies of G\"{o}del-type metrics in various alternative gravity models have been carried out \cite{ourgodel}.

Our aim in this paper is to study the consistency of the G\"{o}del and G\"{o}del-type metrics  within a particular model of covariant galileon theories called cubic galileon gravity model \cite{Babichev:2016fbg, VanAelst:2019kku, Deffayet:2010qz}, that is, the version of galileon theory involving terms up to the third order in the galileon scalar field (as it is known, the most general form of the galileon action includes terms up to the fifth order \cite{Kobayashi:2019hrl}). In \cite{Geng:2017nwv}, the authors have been investigated G\"{o}del-type metrics in Einstein-Horndeski theory, which does not involve the cubic galileon term.

The structure of the paper looks like follows. In section \ref{sec2}, we define the cubic galileon gravity model, in section \ref{sec3}, we check consistency of the usual G\"{o}del metric within it. In section \ref{sec4}, we briefly discuss the main features of G\"{o}del-type metrics. Next, section \ref{sec5} is focused on G\"{o}del-type solutions in the cubic galileon gravity model as well as their causality properties. Finally, in the section \ref{sec6} our conclusions are presented.

\section{The model: Cubic galileon Gravity}
\label{sec2}

The dynamics of the cubic galileon field $\pi$ is described by the action (for a general form of its action see f.e. \cite{Ratt}; in our paper we, for the sake of simplicity, restrict ourselves the particular case $c_4=c_5=0$, i.e. we take into account only terms up to the third order in $\pi$)
\bea
S=\int d^4x\sqrt{-g}\left[\frac{M_{P}^2}{2}R-2\Lambda+\frac{1}{2}\sum_{i=1}^3c_i{\cal L}_i\right]+S_m,
\eea
where $g$ is the metric determinant, $M_{P}$ is the Planck mass, $c_i$'s \footnote{Note that here we are dealing with a more generic cubic galileon gravity than, for example, \cite{Babichev:2016fbg, VanAelst:2019kku}, where $c_{1}$ is taken to be zero. In such cases, the action is invariant under the shift transformation $\pi\rightarrow\pi+const$, which is a subgroup of the enhanced transformation $\pi\rightarrow\pi+b_{\mu}x^{\mu}+const$.} are dimensionless constants, $S_m$ is the action associated with the content of matter and
\bea
{\cal L}_1&=& M^3\pi,\\
{\cal L}_2&=&\left(\nabla\pi\right)^2,\\
{\cal L}_3&=&\frac{1}{M^3}\left(\nabla\pi\right)^2\Box\pi,
\eea
with $M$ being a mass dimension constant and
\bea
\left(\nabla\pi\right)^2&=&g^{\mu\nu}\nabla_\mu\pi\nabla_\nu\pi,\\
\Box\pi&=&g^{\mu\nu}\nabla_\mu\nabla_\nu\pi.
\eea

Varying the action with respect to the metric $g_{\mu\nu}$, the modified Einstein equations are
\bea
G_{\mu\nu}+\Lambda g_{\mu\nu}=M_{P}^{-2}\left[T_{\mu\nu}^{(m)}+T_{\mu\nu}^{(\pi)}\right], \label{FE1}
\eea
where $G_{\mu\nu}$ is the Einstein tensor, $T_{\mu\nu}^{(m)}$ is the energy-momentum tensor associated with the content of matter, $\Lambda$ is the cosmological constant and $T_{\mu\nu}^{(\pi)}$ is the energy-momentum tensor associated with the galileon field which is defined as
\bea
T_{\mu\nu}^{(\pi)}&=&\frac{c_1M^3}{2}g_{\mu\nu}\pi-c_2\left[\nabla_\mu\pi\nabla_\nu\pi-\frac{1}{2}g_{\mu\nu}\left(\nabla\pi\right)^2\right]\nonumber\\
&-&\frac{c_3}{M^3}\left[\nabla_\mu\pi\nabla_\nu\pi\Box\pi-\nabla^\rho\pi\left(\nabla_\mu\pi\nabla_\nu\nabla_\rho\pi+\nabla_\nu\pi\nabla_\mu\nabla_\rho\pi\right)+g_{\mu\nu}\nabla^\alpha\pi\nabla_\alpha\nabla_\beta\pi\nabla^\beta\pi\right].
\eea 
The field equation for the galileon field $\pi$ is given as
\bea
\frac{c_1M^3}{2}-c_2\Box\pi+\frac{c_3}{M^3}\left[-\left(\Box\pi\right)^2+R_{\mu\nu}\nabla^\mu\pi\nabla^\nu\pi+\nabla^\mu\nabla^\nu\pi\nabla_\mu\nabla_\nu\pi\right]=0,\label{FE2}
\eea
where $R_{\mu\nu}$ is the Ricci tensor.

In the next sections,  we employ these equations in order to verify consistency of G\"{o}del and G\"{o}del-type solutions within the framework of the cubic galileon gravity.

\section{G\"{o}del universe in Cubic galileon Gravity}
\label{sec3}
To start with this section, let us outline the main properties of G\"{o}del universe in GR. This metric is a solution of GR with cosmological constant describing a rotating Universe in the presence of a dust source with density $\rho$. Its most remarkable feature is that it presents CTC's, i.e., observers traveling along these closed curves could return to the past and, thereby, violating the causality even though holding valid the locally principles of special relativity. Next, we shall investigate the consistency of the G\"{o}del universe within galileon gravity as well as the causality properties.

Now let us study the field equations (\ref{FE1}) and (\ref{FE2}) for the G\"{o}del universe. Its metric looks like \cite{Godel}:
\bea
ds^2 = a^2 \left(-dt^2 + dx^2 - \frac{1}{2}e^{2x}dy^2 + dz^2 - 2e^xdt \ dy \right),
\eea
where $a$ is an arbitrary number. The relevant tensor quantities associated with this metric are:
\begin{enumerate}
\item Non-zero Christoffel symbols:
\bea
\Gamma^0_{01}=1,\quad\quad\quad \Gamma^0_{12}=\Gamma^1_{02}=\frac{e^x}{2},\quad\quad\quad \Gamma^1_{22}=\frac{e^{2x}}{2},\quad\quad\quad \Gamma^2_{01}=-e^{-x}.
\eea
\item Non-zero Ricci tensor components are
\bea
    R_{00} = 1, \quad\quad\quad R_{02} = R_{20} = e^x, \quad\quad\quad R_{22} = e^{2x}.
\eea
\item The Ricci scalar is
\bea
    R = \frac{1}{a^2}.
\eea
\item Non-zero Einstein tensor components are
\bea
G_{00}=G_{11}=G_{33}=\frac{1}{2},\quad\quad\quad G_{02}=\frac{1}{2}e^x,\quad\quad\quad G_{22}=\frac{3}{4}e^{2x}.
\eea
\end{enumerate}

In order to calculate the energy-momentum tensor associated with the galileon field, let us consider that $\pi=\pi(t)$. Using that
\bea
\nabla_\mu\pi&=&\partial_\mu\pi\\
\nabla_\mu\nabla_\nu\pi&=&\nabla_\mu(\partial_\nu\pi)=\partial_\mu\partial_\nu\pi-\Gamma^\lambda_{\mu\nu}\partial_\lambda\pi,
\eea 
the non-zero components of the energy-momentum tensor are 
\bea
T^{(\pi)}_{00}&=&\frac{c_1M^3}{2}a^2\pi - \frac{c_2}{2}\dot{\pi}^2,\\
T^{(\pi)}_{01}&=&\frac{c_3}{M^3a^2}\dot{\pi}^3,\\
T^{(\pi)}_{02}&=&\frac{c_1M^3}{2}a^2e^x\pi+\frac{c_2}{2}e^x\dot{\pi}^2-\frac{c_3}{M^3a^2}e^x\dot{\pi}^2\ddot{\pi},\\
T^{(\pi)}_{11}&=&-\frac{c_1M^3}{2}a^2\pi-\frac{c_2}{2}\dot{\pi}^2+\frac{c_3}{M^3a^2}\dot{\pi}^2\ddot{\pi},\\
T^{(\pi)}_{22}&=&\frac{c_1M^3}{2}\frac{a^2e^{2x}}{2}\pi+\frac{c_2}{2}\frac{e^{2x}}{2}\dot{\pi}^2-\frac{c_3}{M^3a^2}\frac{e^{2x}}{2}\dot{\pi}^2\ddot{\pi},\\
T^{(\pi)}_{33}&=&-\frac{c_1M^3}{2}a^2\pi-\frac{c_2}{2}\dot{\pi}^2+\frac{c_3}{M^3a^2}\dot{\pi}^2\ddot{\pi}.
\eea

By taking as matter content 
\bea
T^{(m)}_{\mu\nu}=\rho u_\mu u_\nu,
\eea
with $\rho$ being the energy density and $u_\mu=(a, 0, ae^x, 0)$ the 4-velocity, the components of the field equation (\ref{FE1}) take the form
\bea
&(0,0):& \quad \frac{1}{2}=M_{P}^{-2}\left(\rho a^2+\frac{c_1M^3}{2}a^2\pi - \frac{c_2}{2}\dot{\pi}^2 \right)+\Lambda a^2,\\
&(0,1):& \quad 0=M_{P}^{-2}\frac{c_3}{M^3a^2}\dot{\pi}^3,\label{01}\\
&(0,2):& \quad \frac{1}{2}e^x=M_{P}^{-2}\left(\rho a^2e^x+\frac{c_1M^3}{2}a^2e^x\pi+\frac{c_2}{2}e^x\dot{\pi}^2-\frac{c_3}{M^3a^2}e^x\dot{\pi}^2\ddot{\pi}\right)+\Lambda a^2e^x,\\
(1,1)=&(3,3):& \quad \frac{1}{2}=M_{P}^{-2}\left(-\frac{c_1M^3}{2}a^2\pi-\frac{c_2}{2}\dot{\pi}^2+\frac{c_3}{M^3a^2}\dot{\pi}^2\ddot{\pi}\right)-\Lambda a^2,\\
&(2,2):& \quad \frac{3}{4}e^{2x}=M_{P}^{-2}\left(\rho a^2 e^{2x}+\frac{c_1M^3}{2}\frac{a^2e^{2x}}{2}\pi+\frac{c_2}{2}\frac{e^{2x}}{2}\dot{\pi}^2-\frac{c_3}{M^3a^2}\frac{e^{2x}}{2}\dot{\pi}^2\ddot{\pi}\right)+\frac{1}{2}a^2e^{2x}\Lambda.\nonumber\\
\eea
The equation (\ref{01}) leads to
\bea
\dot{\pi}^3=0\quad \Longrightarrow\quad\pi=c,
\eea
where $c$ is an arbitrary constant. The field equation for the field $\pi$, i.e. eq. (\ref{FE2}), provides
\bea
\frac{c_1M^3}{2}=0.
\eea
Thus, in this case the galileon gravity is reduced to the usual Einstein gravity. Therefore, the galileon gravity admits the G\"{o}del solution, that is, the field equations are solved for the condition
\bea
\rho=\frac{M_{P}^2}{a^2}\quad\quad\mathrm{and}\quad\quad \Lambda=-\frac{1}{2a^2}.
\eea
This implies that the CTC's are allowed in the cubic galileon gravity.

 \section{G\"{o}del-type metrics}
\label{sec4}

In this section we briefly discuss the main features of a generalized class of metrics called G\"{o}del-type metrics assuming homogeneity in the space and time (ST-homogeneous ones), in the following we shall discuss the homogeneity conditions. In a wider perspective, the G\"{o}del metric discussed in the former section sets up as a particular example of the class of ST-homogeneous G\"{o}del-type metrics. As shown in \cite{Reb1}, its line element takes the following form in cylindrical coordinates
\begin{equation}\label{type_godel}
    ds^2=-[dt+H(r)d\theta]^2+D^{2}(r)d\theta^{2}+dr^{2}+dz^{2}, 
\end{equation}
where $H(r)$ and $D(r)$ are metric functions depending only on  radius coordinate $r$. Apart from this, the homogeneity conditions in the space-time are achieved by the following relations between metric functions, namely:
\begin{equation}
\begin{split}
&\frac{H^{'}(r)}{D(r)}=2\omega,\\
&\frac{D^{''}(r)}{D(r)}=m^{2},
\end{split}
\label{ST}
\end{equation}
where the prime stands for derivative with respect to the radius coordinate. The pair of constant parameters $(m^2, \omega)$ describes entirely the ST-homogeneity conditions as laid out in Eq. (\ref{ST}). They are restricted to take on values in the range: $-\infty \leq m^2 \leq \infty$ and $\omega\neq0$ (which is physically interpreted as the rotation of the space-time). As remarked in \cite{Reb1}, the ST-homogeneous G\"{o}del-type spaces can be split into three different classes by depending on the sign of $m^2$:
\begin{itemize}
\item \textit{hyperbolic class}: $m^2>0$, $\omega\neq 0$:
\begin{equation}
\begin{split}
&H(r)=\frac{2\omega}{m^2}[\cosh(mr)-1],\\
&D(r)=\frac{1}{m}\sinh(mr),\\
\end{split}
\end{equation}
\item \textit{trigonometric class}: $-\mu^2=m^2<0$, $\omega\neq 0$:
\begin{equation}
\begin{split}
&H(r)=\frac{2\omega}{\mu^2}[1-\cos(\mu r)],\\
&D(r)=\frac{1}{\mu}\sin(\mu r),\\
\end{split}
\label{trigo}
\end{equation}
\item \textit{linear class}: $m^2=0$, $\omega\neq 0$:
\begin{equation}
\begin{split}
&H(r)=\omega r^2,\\
&D(r)=r.\\
\end{split}
\label{linear}
\end{equation}
\end{itemize}
recalling we are getting rid of the degenarate class that corresponds to $\omega=0$. Noteworthy that the G\"{o}del metric ($m^2 =2\omega^2$) belongs to the hyperbolic class. Another important feature of the ST-homogeneous G\"{o}del-type spaces concern to the isometry group, for example: the class $m^2=4\omega^2$ admits the larger isometric group, $G_7$ \cite{Reb1}, whilst for $m^2<4\omega^2$ admits $G_5$ as the isometry group.

The ST-homogeneous G\"{o}del-type spaces present Closed Time-like Curves (CTC's) which are circles $C=\lbrace(t,r,\theta,z); \, t, r, z= \mbox{const}, \theta \in [0, 2\pi]\rbrace$, defined in a region limited by the range ($r_1 <r<r_2$), where $G(r)=D^2 (r)-H^2 (r)$ becomes negative within this range. It is interesting to note that there is not CTC's for the hyperbolic class corresponding to $m^2\geq 4\omega^2$, otherwise, it does. Hence, for the hyperbolic class with range of parameters $0<m^2<4\omega^2$ there exists CTC's inside the region corresponding to $r>r_{c}$, where $r_{c}$ is the critical radius (limiting radius separating the causal and non-causal regions), its explicit form is given by
\begin{equation}
\sinh^2\bigg(\frac{m r_{c}}{2}\bigg)=\bigg(\frac{4\omega^2}{m^2}-1 \bigg)^{-1}.
\label{rc}
\end{equation}  
Similarly, the linear and trigonometric classes also exhibit CTCs. Both cases display a non-causal region, namely: for the linear one, this region is hit for $r>r_{c}$ and the critical radius $r_{c}=\frac{1}{\omega}$. In the trigonometric case, the situation is more subtle since there exists an infinite set of alternating non-causal and causal regions (see for example for an explicit form of $r_{c}$). In the next section, we shall check the consistency of ST-homogeneous G\"{o}del-type metrics and also their causality properties inside the galileon gravity.

\section{G\"{o}del-type solution in Cubic galileon Gravity}
\label{sec5}

 We now focus our attention on the study of G\"{o}del-type metrics in cubic galileon gravity. In order to proceed any further, let us define a local set of tetrad basis $\theta^{A}=e^A\ _{\mu}\ dx^\mu$, the reason is only to make calculations simpler as we will see later. In particular, a good choice for the tetrad basis looks like:
    \bea
    \theta^{(0)} = dt + H(r)d\phi, \quad \theta^{(1)} = dr, \quad \theta^{(2)} = D(r)d\phi, \quad \theta^{(3)} &= dz, 
			\label{tetrad}
\eea
where we have adopted capital Latin letters to label tetrad indices. Thus, the line element takes the form   
\begin{equation}
    ds^2 = \eta_{AB} \theta^A \theta^B = -(\theta^{(0)})^2 + (\theta^{(1)})^2 + (\theta^{(2)})^2 + (\theta^{(3)})^2, \label{frame}
\end{equation}
where $\eta_{AB}$ is the Minkowski metric.

Thus the field equations in the tetrad basis (\ref{tetrad}) becomes
\bea
G_{AB}+\Lambda g_{AB}=M_{P}^{-2}\left[T_{AB}^{(m)}+T_{AB}^{(\pi)}\right], \label{FE3}
\eea
the dynamical galileon equation
\bea
\frac{c_1M^3}{2}-c_2\Box\pi+\frac{c_3}{M^3}\left[-\left(\Box\pi\right)^2+R_{AB}\nabla^A\pi\nabla^B\pi+\nabla^A\nabla^B\pi\nabla_A\nabla_B\pi\right]=0,\label{FE4}
\eea
where
\bea
G_{AB}=e^\mu_A e^\nu_B G_{\mu\nu}, \quad T_{AB}=e^\mu_A e^\nu_B T_{\mu\nu}, \quad g_{AB}=e^\mu_A e^\nu_B g_{\mu\nu}, \quad \nabla_A=e^\nu_A\nabla_\nu.
\eea
with $e^\mu\ _{B}$ being the inverse of $e^A\ _{\mu}$ and then satisfying the following condition  $e^A\ _{\mu}e^\mu\ _{B}=\delta^{A}_{B}$.

By taking $\pi=\pi(t)$ the d'Alembertian operator takes the form
\bea
\Box\pi=\left(\frac{D^2-H^2}{D^2}\right)\ddot{\pi}.
\eea
The non-zero components of the energy-momentum tensor associated to the galileon field are
\bea
T^{(\pi)}_{(0)(0)}&=&-\frac{c_1M^3}{2}\pi-\frac{c_2}{2}\left(1+\frac{H^2}{D^2}\right)\dot{\pi}^2-\frac{c_3}{M^3}\frac{H^2}{D^2}\left(1-\frac{H^2}{D^2}\right)\dot{\pi}^2\ddot{\pi},\\
T^{(\pi)}_{(0)(1)}&=&-\frac{c_3}{M^3}\frac{H}{D}\left(2\omega-\frac{HD'}{D^2}\right)\dot{\pi}^3,\\
T^{(\pi)}_{(0)(2)}&=&-c_2\frac{H}{D}\dot{\pi}^2-\frac{c_3}{M^3}\frac{H}{D}\left(1-\frac{H^2}{D^2}\right)\dot{\pi}^2\ddot{\pi},\\
T^{(\pi)}_{(1)(1)}&=&\frac{c_1M^3}{2}\pi-\frac{c_2}{2}\left(1-\frac{H^2}{D^2}\right)\dot{\pi}^2-\frac{c_3}{M^3}\left(1-\frac{H^2}{D^2}\right)^2\dot{\pi}^2\ddot{\pi},\\
T^{(\pi)}_{(1)(2)}&=&-\frac{c_3}{M^3}\frac{H}{D}\left(2\omega-\frac{HD'}{D^2}\right)\dot{\pi}^3,\\
T^{(\pi)}_{(2)(2)}&=&\frac{c_1M^3}{2}\pi-\frac{c_2}{2}\left(1+\frac{H^2}{D^2}\right)\dot{\pi}^2-\frac{c_3}{M^3}\left(1-\frac{H^2}{D^2}\right)\dot{\pi}^2\ddot{\pi},\\
T^{(\pi)}_{(3)(3)}&=&\frac{c_1M^3}{2}\pi-\frac{c_2}{2}\left(1-\frac{H^2}{D^2}\right)\dot{\pi}^2-\frac{c_3}{M^3}\left(1-\frac{H^2}{D^2}\right)^2\dot{\pi}^2\ddot{\pi}.
\eea

To write the field equations the following quantities are necessary. The non-zero components of Einstein tensor, in the tetrad basis (\ref{tetrad}), are
\bea
G_{(0)(0)}=3\omega^2-m^2,\quad\quad G_{(1)(1)}=G_{(2)(2)}=\omega^2, \quad\quad G_{(3)(3)}=m^2-\omega^2.
\eea
It remains only to fix the matter content as the last ingredient to complete the field equations. The well-motivated matter sources for G\"{o}del-type metrics have been worked out in \cite{Reb} where they were shown to be presented by a perfect fluid, a scalar field and an electromagnetic field. Here, we will pick a perfect fluid and an electromagnetic field as matter sources only. Let us get started by describing the perfect fluid whose the energy-momentum tensor in the tetrad basis (\ref{tetrad}) is given by
\bea
T^{(pf)}_{AB}= (\rho+p)u_Au_B-p\eta_{AB}\label{EMT},
\eea
where $u^{A}=e^{A}_{0}$ is the 4-velocity of the fluid defining in the comoving frame and $\rho$ and $p$ are the density and pressure of the fluid, respectively . The non-zero components are
\bea
T^{(pf)}_{(0)(0)}=\rho, \quad\quad\quad T^{(pf)}_{(1)(1)}=T^{(pf)}_{(2)(2)}=T^{(pf)}_{(3)(3)}=p.
\eea
Regarding the electromagnetic field $F_{AB}$, we assume $F_{AB}$ in such a way that the electric and magnetic fields lie in $z$-direction in agreement to \cite{Reb}. In this case, the non-zero components of the energy-momentum tensor are
\bea
T^{(ef)}_{(0)(0)}=T^{(ef)}_{(1)(1)}=T^{(ef)}_{(2)(2)}=\frac{e^2}{2}, \quad\quad\quad T^{(ef)}_{(3)(3)}=-\frac{e^2}{2},
\eea
where $e$ is the amplitude of the electromagnetic field. Therefore, the energy-momentum tensor of the matter sources is
\bea
 T_{AB}^{(m)}=T_{AB}^{(pf)}+T_{AB}^{(ef)},
\label{Tm}
\eea
that, as a result of the symmetries of the space-time, it is completely diagonal and their components are constants. Accordingly, the field equations, eq. (\ref{FE3}), in the tetrad basis, look like
\bea
3\omega^2-m^2-\Lambda&=&M_{P}^{-2}\rho+M_{P}^{-2}\frac{e^2}{2}+M_{P}^{-2}\Biggl[-\frac{c_1M^3}{2}\pi-\frac{c_2}{2}\left(1+\frac{H^2}{D^2}\right)\dot{\pi}^2-\nonumber\\ &-&\frac{c_3}{M^3}\frac{H^2}{D^2}\left(1-\frac{H^2}{D^2}\right)\dot{\pi}^2\ddot{\pi}\Biggl],\\
\omega^2+\Lambda&=&M_{P}^{-2}p+M_{P}^{-2}\frac{e^2}{2}+M_{P}^{-2}\left[\frac{c_1M^3}{2}\pi-\frac{c_2}{2}\left(1-\frac{H^2}{D^2}\right)\dot{\pi}^2-\right.\nonumber\\ &-&\left. \frac{c_3}{M^3}\left(1-\frac{H^2}{D^2}\right)^2\dot{\pi}^2\ddot{\pi}\right],\\
\omega^2+\Lambda&=&M_{P}^{-2}p+M_{P}^{-2}\frac{e^2}{2}+M_{P}^{-2}\left[\frac{c_1M^3}{2}\pi-\frac{c_2}{2}\left(1+\frac{H^2}{D^2}\right)\dot{\pi}^2-\right.\nonumber\\ &-&\left. \frac{c_3}{M^3}\left(1-\frac{H^2}{D^2}\right)\dot{\pi}^2\ddot{\pi}\right],\\
m^2-\omega^2+\Lambda&=&M_{P}^{-2}p-M_{P}^{-2}\frac{e^2}{2}+M_{P}^{-2}\left[\frac{c_1M^3}{2}\pi-\frac{c_2}{2}\left(1-\frac{H^2}{D^2}\right)\dot{\pi}^2-\right.\nonumber\\ &-& \left.\frac{c_3}{M^3}\left(1-\frac{H^2}{D^2}\right)^2\dot{\pi}^2\ddot{\pi}\right].
\eea
These are the diagonal components $(0,0), (1,1), (2,2)$ and $(3,3)$, respectively. The off-diagonal components $(0,1), (0,2)$ and $(1, 2)$, are, respectively
\bea
-\frac{c_3}{M^3}\frac{H}{D}\left(2\omega-\frac{HD'}{D^2}\right)\dot{\pi}^3&=&0,\\
-c_2\frac{H}{D}\dot{\pi}^2-\frac{c_3}{M^3}\frac{H}{D}\left(1-\frac{H^2}{D^2}\right)\dot{\pi}^2\ddot{\pi}&=&0,\\
-c_2\frac{H}{D}\dot{\pi}^2-\frac{c_3}{M^3}\frac{H}{D}\left(1-\frac{H^2}{D^2}\right)\dot{\pi}^2\ddot{\pi}&=&0.
\eea
A direct inspection in these off-diagonal components imply that the field $\pi$ is a constant. In addition, the field equation for the galileon field, eq. (\ref{FE4}), leads to $\frac{c_1M^3}{2}=0$. Therefore, in this case the G\"{o}del-type solutions in galileon gravity reduce to the GR solutions. For example, in the absence of electromagnetic field, the condition $m^2=2\omega^2$ is obtained and CTC's are allowed \cite{Godel}. As a consequence, in this case with $\pi=\pi(t)$, the violation of causality is permitted.

The question as to whether another dependence on the galileon field could generate G\"{o}del-type causal solutions naturally arises at this point. In order to do such an investigation, let us consider $\pi=\pi(z)$, which is a suggestive choice, since the non-trivial dependence of the galileon with the rotating axis ($z-$axis) might lead to dynamical non-trivial effects and affect the causality properties correspondingly.

For this case, the d'Alembertian operator acts on the field as
\bea
\Box\pi(z)=-\pi''(z),
\eea
where the prime denotes the derivative with respect to $z$. The non-zero components of the energy-momentum tensor are
\bea
T^{(\pi)}_{(0)(0)}&=&-\frac{c_1M^3}{2}\pi-\frac{c_2}{2}\pi'^2 +\frac{c_3}{M^3}\pi'^2\pi'',\\
T^{(\pi)}_{(1)(1)}&=&T^{(\pi)}_{(2)(2)}=\frac{c_1M^3}{2}\pi+\frac{c_2}{2}\pi'^2 -\frac{c_3}{M^3}\pi'^2\pi'',\\
T^{(\pi)}_{(3)(3)}&=&\frac{c_1M^3}{2}\pi-\frac{c_2}{2}\pi'^2.
\eea
Then the non-zero components of the field equation are
\bea
3\omega^2-m^2-\Lambda&=&M_{P}^{-2}\rho+M_{P}^{-2}\frac{e^2}{2} +M_{P}^{-2}\Biggl[-\frac{c_1M^3}{2}\pi-\frac{c_2}{2}\pi'^2 +\frac{c_3}{M^3}\pi'^2\pi''\Biggl],\label{1}\\
\omega^2+\Lambda&=&M_{P}^{-2}p+M_{P}^{-2}\frac{e^2}{2}+M_{P}^{-2}\left[\frac{c_1M^3}{2}\pi+\frac{c_2}{2}\pi'^2 -\frac{c_3}{M^3}\pi'^2\pi''\right],\label{2}\\
m^2-\omega^2+\Lambda&=&M_{P}^{-2}p-M_{P}^{-2}\frac{e^2}{2}+M_{P}^{-2}\left[\frac{c_1M^3}{2}\pi-\frac{c_2}{2}\pi'^2\right],\label{3}
\eea
where Eq. (\ref{Tm}) has been used.

The field equation for the galileon, eq. (\ref{FE4}), reduces to
\bea
\frac{c_1M^3}{2}-c_2\pi''=0,
\eea
whose solution is 
\bea
\pi(z)=B+Az+\frac{M^3c_1}{4c_2}z^2,\label{pi}
\eea
where $A$ and $B$ are integration constants. Now, plugging Eq. (\ref{pi}) into Eqs. (\ref{1}), (\ref{2}) and (\ref{3}), we arrive at
\bea
\nonumber 3\omega^2-m^2-\Lambda&=&M_{P}^{-2}\rho+M_{P}^{-2}\frac{e^2}{2}-\frac{z^2}{4}\left(\frac{c_{1}^{2}M^6}{M_{p}^{2}c_2}-\frac{1}{2}\frac{c_{3}c_{1}^{3}M^6}{M_{p}^{2}c_{2}^{3}}\right)-z\left(\frac{c_{1}M^{3}A}{M_{p}^{2}}-\frac{1}{2}\frac{c_{3}c_{1}^{2}M^{3}A}{M_{p}^{2}c_{2}^{2}}\right)-\\
&-&\frac{1}{2}\frac{c_{2}A^2}{M_{p}^{2}}+\frac{1}{2}\frac{c_{3}c_{1}A^2}{M_{p}^{2}c_{2}}-\frac{1}{2}\frac{c_{1}M^{3}B}{M_{p}^{2}},\label{11}\\
\nonumber \omega^2+\Lambda&=&M_{P}^{-2}p+M_{P}^{-2}\frac{e^2}{2}+\frac{z^2}{4}\left(\frac{c_{1}^{2}M^6}{M_{p}^{2}c_2}-\frac{1}{2}\frac{c_{3}c_{1}^{3}M^6}{M_{p}^{2}c_{2}^{3}}\right)+z\left(\frac{c_{1}M^{3}A}{M_{p}^{2}}-\frac{1}{2}\frac{c_{3}c_{1}^{2}M^{3}A}{M_{p}^{2}c_{2}^{2}}\right)+\\
&+&\frac{1}{2}\frac{c_{2}A^2}{M_{p}^{2}}-\frac{1}{2}\frac{c_{3}c_{1}A^2}{M_{p}^{2}c_{2}}+\frac{1}{2}\frac{c_{1}M^{3}B}{M_{p}^{2}},\label{22}\\
m^2-\omega^2+\Lambda&=&M_{P}^{-2}p-M_{P}^{-2}\frac{e^2}{2}-\frac{1}{2}\frac{c_{2}A^2}{M_{p}^{2}}+\frac{1}{2}\frac{c_{1}M^{3}B}{M_{p}^{2}}.\label{33}
\eea
Note that the {\it r.h.s.} of the former equations explicitly depend on the coordinate $z$ while the {\it l.h.s.} does not, thus for the sake of consistency one requires constraints among the $c_{i}$'s coupling constants, namely: either $2c_{2}^{2}=c_{1}c_{3}$ or $c_{1}=0$, thereby precluding $z$-coordinate dependency.

	\begin{itemize}
		\item First case: $2c_{2}^{2}=c_{1}c_{3}$.
	\end{itemize}
	 In this situation, the aforementioned equations reduce to
\bea
3\omega^2-m^2-\Lambda&=&M_{P}^{-2}\rho+M_{P}^{-2}\frac{e^2}{2}+\frac{1}{2}\frac{c_{2}A^2}{M_{p}^{2}}-\frac{1}{2}\frac{c_{1}M^{3}B}{M_{p}^{2}},\label{111}\\
\omega^2+\Lambda&=&M_{P}^{-2}p+M_{P}^{-2}\frac{e^2}{2}-\frac{1}{2}\frac{c_{2}A^2}{M_{p}^{2}}+\frac{1}{2}\frac{c_{1}M^{3}B}{M_{p}^{2}},\label{222}\\
m^2-\omega^2+\Lambda&=&M_{P}^{-2}p-M_{P}^{-2}\frac{e^2}{2}-\frac{1}{2}\frac{c_{2}A^2}{M_{p}^{2}}+\frac{1}{2}\frac{c_{1}M^{3}B}{M_{p}^{2}}.\label{333}
\eea
 Now, subtracting Eq.(\ref{222}) to (\ref{333}), we have
\bea
m^2-2\omega^2&=&-\frac{e^{2}}{M_{p}^{2}},\label{first}
\eea
which is a striking relation linking up the metric parameters with the matter content. It shows that $m^2$ is bounded on the top, i.e., $m^2\leq 2\omega^2$, as a result, no completely causal solution can be found. On the other hand, there is no lower bound in $m^2$, thus all three classes of G\"{o}del-type metrics are achieved. The hyperbolic class is covered for $m^2=2\omega^2-\frac{e^{2}}{M_{p}^{2}}>0$, the linear class is covered for $\omega^2=\frac{e^{2}}{2 M_{p}^{2}}$ while the trigonometric class is obtained for $\mu^2=\frac{e^{2}}{M_{p}^{2}}-2\omega^2>0$. In particular, the G\"{o}del solution is reached in the absence of electromagnetic field as can be seen from Eq. (\ref{first}).   

Combining Eqs. (\ref{222}, \ref{333}, \ref{first}) we find the cosmological constant to be
\bea 
\Lambda=M_{p}^{-2}\left(\frac{p}{2}-\frac{\rho}{2}+\frac{e^2}{2}-c_{2}A^2 +c_{1}M^{3}B\right).
\eea
Note that the cosmological constant depends on the parameters of the model and the matter content as well. Therefore, this relation is more generic than in GR where the cosmological constant is entirely determined by the matter content while in cubic galileon theory is not.  

\begin{itemize}
		\item Second case: $c_{1}=0$.
	\end{itemize}
		
Now, the set of gravitational field equations reduce to
\bea
3\omega^2-m^2-\Lambda&=&M_{P}^{-2}\rho+M_{P}^{-2}\frac{e^2}{2}-\frac{1}{2}\frac{c_{2}A^2}{M_{p}^{2}},\label{1111}\\
 \omega^2+\Lambda&=&M_{P}^{-2}p+M_{P}^{-2}\frac{e^2}{2}+\frac{1}{2}\frac{c_{2}A^2}{M_{p}^{2}},\label{2222}\\
m^2-\omega^2+\Lambda&=&M_{P}^{-2}p-M_{P}^{-2}\frac{e^2}{2}-\frac{1}{2}\frac{c_{2}A^2}{M_{p}^{2}}.\label{3333}
\eea
Notice that, in this case, the cubic term plays an irrelevant role in the above equations. From Eqs.(\ref{2222}) and (\ref{3333}) we find 
\bea
m^2 -2\omega^2 =-\frac{1}{M_{p}^{2}}\left(e^2 +c_{2}A^2\right).
\label{xdx}
\eea
Since we are looking for possible completely causal solutions we must demand $c_{2}=-|c_{2}|$ and also $|c_{2}|\geq e^2$ to find the condition
\bea
 m^2\geq 2\omega^2,
\eea
which displays a lower bound for $m^2$. Then, only solutions within the hyperbolic class are admissible. The G\"{o}del metric is arrived at by imposing $e^2=|c_{2}|A^2$ as can be seen from Eq.(\ref{xdx}).

 To proceed further, by summing Eqs.(\ref{1111}) and (\ref{2222}) we arrive at
\bea
m^2-4\omega^2=-M_{p}^{-2}(p+\rho+e^2)\leq 0,
\eea
that shows that neither parameters of the model nor the galileon field  are involved. Remarkably, this relation is exactly the same to GR even in the presence of the galileon. Hence, the admissible solutions are restricted to the range $2\omega^2\leq m^2\leq 4\omega^2$.

This result leads to some possible causal G\"{o}del-type solutions. (i) For an empty universe, that is, $e=\rho=p=0$, with the galileon field of the form $\pi(z)=B+Az$, the condition $m^2=4\omega^2$ is obtained. (ii) If the universe is filled with an exotic fluid (dark energy) such that $p=-\rho$, in the absence of electromagnetic field and galileon $\pi(z)=B+Az$, the causal condition is permitted.

The cosmological constant is readily obtained from Eqs.(\ref{1111},\ref{2222},\ref{3333}) and it reads
\bea
\Lambda=M_{p}^{-2}\left(\frac{p}{2}-\frac{\rho}{2}+\frac{e^2}{2}-|c_{2}|A^2\right).
\eea
As expected, similar to the former case, the cosmological constant depends on the parameters of the model and also the Galileon field form. Formally, such a solution is identical to GR plus a scalar field for the particular case: $|c_{2}|=1$, \cite{Reb}. Of course, it happens as a result of the contribution of the cubic term vanishes in the field equations, when $c_{1}=0$.

\section{Summary and Conclusions}
\label{sec6}

We discussed the G\"{o}del and G\"{o}del-type metrics within the cubic galileon gravity with the galileon field taking the following forms: $\pi=\pi(t)$ and $\pi=\pi(z)$. In addition, in the presence of the matter sources, namely: a perfect fluid and an electromagnetic field. In this outlook, we have succeed to generate G\"{o}del-type solutions having no analogy with GR only for $\pi=\pi(z)$, in the case $\pi=\pi(t)$ the solution is trivial and reduces to GR identically. Regarding the non-trivial case $\pi=\pi(z)$, the coupling constant must satisfy two constraint for consistency of the field equations.

The first case occurs when the coupling constants satisfy the constraint: $2c_{2}^{2}=c_1 c_{3}$. In this case we remarked that the solutions have a lower upper bound giving by $m^{2}\leq 2\omega^2$, where the equality is accomplished in the absence of electromagnetic field and then corresponds to the G\"{o}del solution in the cubic galileon gravity. In any situation, one cannot find completely causal solution ($m^2\geq 4\omega^2$). Furthermore, solving the dynamical equation for the galileon
we found that its form is quadratic in $z$-coordinate which is completely different from the results obtained in other models \cite{ourgodel}.

The second case occurs for $c_{1}=0$, such a situation is fairly discussed in the literature \cite{Babichev:2016fbg,VanAelst:2019kku,Deffayet:2010qz} in other contexts. In a distinguishing way to the former case, now, we shown that, apart from a lower upper bound for $m^2$, there exists a lower bound as well, i.e., $2\omega^2\leq m^2\leq 4\omega^2$. This result is only reached by requiring $c_{2}=-|c_{2}|$ and $c_{3}$ can be generic. The completely causal solution was found corresponding to either the vacuum solution or for an exotic fluid $p=-\rho$ with no electromagnetic field. Apart from this, the galileon must be linear in $z$-coordinate, albeit this result is not surprising since the cubic galileon theory with $c_{1}=0$ is invariant under shift transformation $\pi\rightarrow \pi+c$ representing themselves as a particular case of "galilean"-like transformations $\pi\to\pi+b_{\mu}x^{\mu}+c$ mentioned in the introduction.

The natural continuation of this study could consist in consideration of more generic galileon gravity involving quartic and quintic terms. Besides of this, clearly, an important direction of study within the framework of the galileon gravity could consist in treating other interesting metrics, such as various black hole and wormhole solutions. We plan to do these studies in forthcoming papers.

\section*{Acknowledgments}

The work by A. Yu. P. is partially supported by the CNPq project 301562/2019-9. This work by A. F. S. is supported by CNPq projects 308611/2017-9 and 430194/2018-8. P. J. P. would like to thank the Brazilian agency CAPES for the financial support.

\end{document}